# Single photonics at telecom wavelengths using nanowire superconducting detectors

**C. Zinoni, B. Alloing, L.H. Li, F. Marsili and A. Fiore**

*Ecole Polytechnique Fédérale de Lausanne, Institute of Photonics and Quantum Electronics*
*Station 3, CH-1015 Lausanne, Switzerland*

**L. Lunghi, A. Gerardino**

*Institute of Photonics and Nanotechnology, CNR, via del Cineto Romano 42, 00156 Roma, Italy*

**Yu. B. Vakhtomin, K. V. Smirnov, G.N. Gol'tsman**

*Moscow State Pedagogical University, 29 M. Pirogovskaya str., 119891 Moscow, Russia*

**Single photonic applications – such as quantum key distribution – rely on the transmission of single photons, and require the ultimate sensitivity that an optical detector can achieve. Single-photon detectors must convert the energy of an optical pulse containing a single photon into a measurable electrical signal. We report on fiber-coupled superconducting single-photon detectors (SSPDs) with specifications that exceed those of avalanche photodiodes (APDs), operating at telecommunication wavelength, in sensitivity, temporal resolution and repetition frequency. The improved performance is demonstrated by measuring the intensity correlation function $g^{(2)}(\tau)$ of single-photon states at 1300nm produced by single semiconductor quantum dots (QDs).**

While silicon-based avalanche photodiodes (APDs) can be used as single-photon detectors in the visible wavelength range, a mature technology with single-photon sensitivity is still lacking at telecom wavelengths (1310-1550 nm). The poor performance of InGaAs APDs barely satisfies the requirements for secure quantum key distribution, limiting the maximum distance and sifted key rate. More generally, progress in the development of systems based on processing and measuring photonic quantum bits at these wavelengths is hindered by the lack of adequate detectors. As an example, while single-photon sources based on semiconductor quantum dots (QDs) have been extensively characterized at $\lambda<1000$ nm, only few clear demonstrations of single-photon emission have been reported at 1300 nm[1,2]. The recent development of single-photon detectors based on NbN superconducting nanostructures[3], promise orders-of-magnitude improvement over InGaAs APDs in sensitivity, dark count, jitter and repetition frequency. The detection principle of the SSPDs is based on the local inhibition of superconductivity in a current-biased ultra-thin superconducting nanowire due to the absorption of a light quantum[4]. As the superconducting energy gap is about 2meV for NbN, an absorbed near-infrared photon supplies a sufficient amount of energy to brake a Cooper pair and promote an electron to a highly excited state. During the relaxation process secondary carriers are efficiently generated and the local electron effective temperature exceeds the superconducting critical temperature, thus creating a 'hot spot'. As the supercurrent avoids the hot spot and accumulates towards the edge of the nanowire, the critical current density may be reached inducing a transition to a resistive state and a subsequent pulse in the external circuit. In the following we report on fiber-coupled SSPDs with sensitivity several orders of magnitude larger than InGaAs APDs, and demonstrate their application to single-photon measurements at telecom wavelengths.

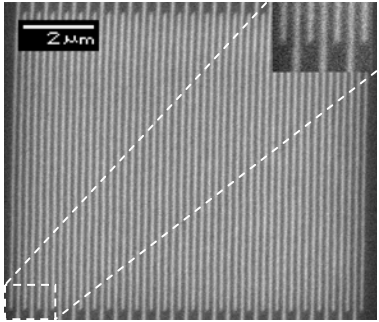

FIG. 1. SEM image of the 10x10mm NbN meander composed of 100nm wide and 3.5nm thick wires.

The SSPDs used in this work consist of 10x10μm$^2$ meanders (fig1) made of a 100nm wide, 3.5nm thick NbN nanowire. Two SSPD chips are mounted on a micromechanical support and aligned to a pair of single mode fibers, the system is then cooled down to 2.3K. The fabrication technology is described in Ref.[5]. The dependence of the detection efficiency (DE) as a function of bias current is shown along with the dark count rate (DCR) in Figure 2a for the best SSPD. The DE for the SSPD refers to the percentage of single-photon pulses incident on the detector that will produce an output signal. Both DE and dark counts increase with increasing bias current. At a bias current of $I_B$=22 μA an efficiency DE=4.8% and dark count rate DCR=13Hz are measured. The other SSPD had lower efficiency (2.5%) for the same DCR. The corresponding DE/DCR ratios are over a factor of 10 higher than previous reports for fiber-coupled SSPDs [6,7]. The tail in the DCR, measured below a bias current of 25μA (Figure 2a), is attributed to the room temperature background radiation entering the system, which can be suppressed by a cold filter.

A direct comparison with APDs is complicated by the fact that InGaAs APDs must be operated in gated mode, so that only a dark count probability (DCP) per gate (1ns) can be defined, which is a function of the gate width. In order to provide an operative comparison of SSPDs and APDs, for autocorrelation experiments on QDs, we define a signal to noise ratio normalized to a time window of 1ns (see Methods for details). In Figure 2b we plot the SNR for the SSPD and the APD as a function of the detection efficiencies. The APD data point corresponds to the optimized working regime used for antibunching measurements in Ref.[2]. The SSPD display an improvement of several orders of magnitude improvement in the SNRs.

Figure 3 reports the temporal dynamics of a laser diode pulse as measured directly using a sampling oscilloscope and by time-correlated single-photon counting (TCSPC) using the SSPD and the APD. From the jitter characteristics of the correlation card, we estimate the time resolution of the SSPD and APD at 150ps and 400ps respectively. Lower jitter values (18ps have been reported[8]) could be achieved by improving the amplification electronics. We note that the SSPD response follows a Gaussian distribution, while the APD presents an asymmetric profile, which depends on the count rate and limits their application for TCSPC experiments.

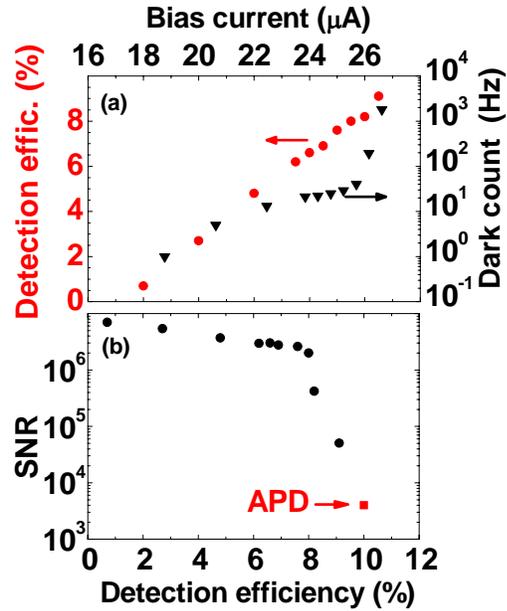

FIG. 2. (a) SSPD detection efficiency and dark count rate plotted as a function of bias current at 2.3K. (b) Comparison between SSPD and APD performance.

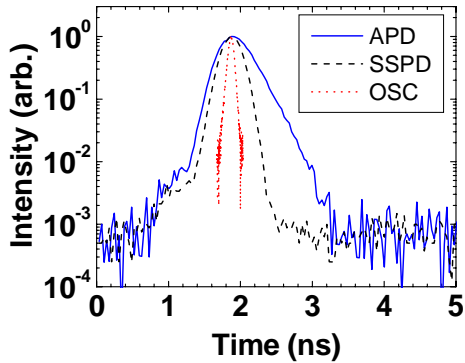

FIG. 3. Measurement of the jitter of the detectors. Dotted line: laser pulse (FWHM=88ps) measured on a sampling oscilloscope (OSC) with 24GHz bandwidth. Continuous line: temporal response of the APD (FWHM=480ps). Dashed line: temporal response of the SSPD (FWHM=320ps).

In order to validate the application of SSPDs to the characterization of non-classical light states at telecom wavelengths, we measured the correlations between single photon pulses produced by a single InAs/GaAs quantum dot emitting at 1300 nm (see Methods). The characterization of the photoluminescence (PL) dynamics and single-photon emission from this type of QDs using APDs was reported in Ref. [2]. The PL spectrum from a single QD (Fig. 4a) is dominated by two lines corresponding to the emission from exciton (X) and biexciton (BX) states. As a first test we demonstrate single photon emission from the X line under non resonant pulsed excitation at 80 MHz. We remark that this repetition frequency is unachievable with APDs, which are limited to a few MHz. The coincidence histogram (Figure 4) is characterized by periodic peaks separated by the laser repetition frequency except for zero delays; this is the signature of a single photon emitter under pulsed excitation. The peaks are well fitted by Gaussian time distributions with an offset of 3 coincidences and a FWHM = 2.2ns that corresponds to twice the X lifetime[2]. Normalizing the counts in 2.2ns windows we calculate a $g^{(2)}(0)=0.18$. This value is lower than the result ($g^{(2)}(0)=0.38$) obtained under similar conditions using two InGaAs APDs[2] due to the 10-fold increase of the SNR in this experiment. The coincidences between the peaks are due to uncorrelated light entering the system and detector dark counts, thus increasing the $g^{(2)}(0)$.

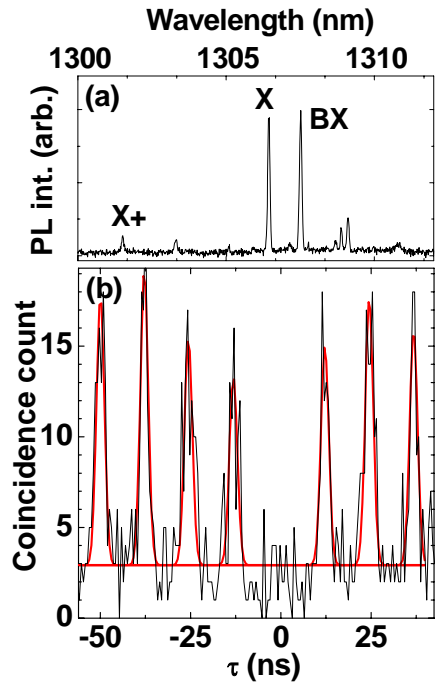

FIG. 4. (a) PL spectrum, at 10K, from a single quantum dot showing emission of the exciton (X), biexciton (BX) and charged exciton (X+). (b) Photon autocorrelation measurement on the X line under pulsed excitation obtained with an integration time of 1.8hrs and a time bin of 560ps. Solid (red) line is a sum of Gaussians with FWHM=2.2ns and an offset of 3 coincidences.

While these pulsed measurements allow an estimation of $g^{(2)}(0)$, which is important for applications that use single-photon sources, many important physical properties of the system are hidden in the correlation function $g^{(2)}(\tau)$ *for any delay* $\tau$[9,10,11]. The measurement of the $g^{(2)}(\tau)$ with APDs is exceedingly difficult since the use of relatively long (>10ns) gates would drive the experimental SNR below unity. The $g^{(2)}(\tau)$ was measured on the positively charged exciton emission (inset fig 5b) under cw excitation by pumping resonantly in the excited state of the trion. The resulting histogram, measured for a pump power of ~0.2mW, is shown in Fig. 5a for long delays (~0.5μs). For time delays between 3ns and 200ns an increase of the correlation function is observed: this bunching behavior, already studied for short-wavelength QDs[12], shows that after emission of a photon from the positive trion the QD remains charged allowing the re-excitation of the charged exciton state. For short time delays $\tau \leq 3$ns, (see Fig. 5(b)), an antibunching dip is observed, confirming the sub-Poissonian statistics of the light emitted by the trion line. The bunching and antibunching behavior can be modeled in a three level system with the following expression[13]: $g^{(2)}(\tau) = 1-(1+a)*\exp(-\tau/\tau_1)+a*\exp(-\tau/\tau_2)$. Taking into account the setup resolution and dark counts[9] the fit yields $g^{(2)}(0)=0.18\pm0.02$ (in agreement with the pulsed-mode results) and a bunching time constant of 171ns.

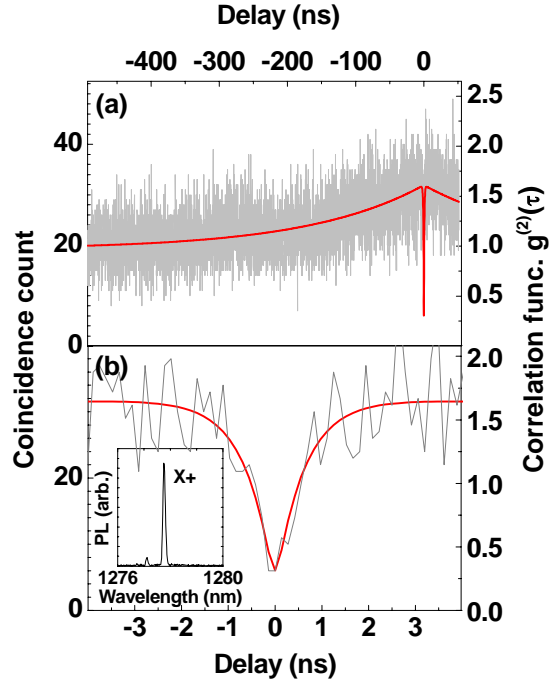

FIG. 5. (a) Measurement of $g^{(2)}(t)$ under resonant excitation, with a time bin of 139ps. (b) Blow-up at short time delays demonstrating antibunching. Inset: PL spectrum of the positive trion under resonant excitation (T=10K). Solid (red) line in (a) and (b) is the fit of the correlation function $g^{(2)}(\ )$.

Photon correlation experiments are a key resource in quantum optics, for example for the characterization of thermal light sources[14], the measurement of spatial coherence[15], and in quantum coherence optical tomography[16]. We thus expect that the availability of these SSPD systems will significantly improve the range of application of quantum technologies in the telecommunication wavelength range. We anticipate that the dark count can be reduced by more than one order of magnitude with appropriate filtering of the room temperature background radiation and even higher efficiencies can be achieved by integrating optical microcavities with the SSPD - up to 50% was demonstrated for free-space coupling[17]. SSPDs thus have a considerable potential impact in photonics and quantum optics.

## Methods

Two single-mode fibers were separately aligned to the SSPD chips using micromechanical precision holders. Input fiber connectors and output SMA cables are mounted at room temperature on the flange of the cryogenic insert. The insert is immersed in a liquid helium bath and the detectors are cooled down to 2.3K by reducing the He vapor pressure. The high frequency components of the detector output are fed to a 60 dB-gain amplifier with a 2.5 GHz bandwidth. In the autocorrelation experiments the SSPDs are biased through bias-Ts to give a DCR of 10-30Hz. The efficiency of the fiber coupled detector is measured using a gain-switched diode laser at 1300nm and a calibrated APD as a reference. In fig.2b the SNR=DE/DCP values for the SSPD are deduced from the data presented in fig.2a. For the APD we used the same definition of SNR: the DCP (for a 1ns gate) was extrapolated from the dark count rate (30Hz) measured during in a 300ps optical active window and at a repetition frequency of 4MHz[2].

The antibunching measurements were made on InAs QDs. The low-density (~2dots/µm$^2$) ensemble of QDs was grown by molecular beam epitaxy at the center of a GaAs/AlAs planar microcavity, as described in Ref. [[18]]. Circular mesas of 1.5 µm diameter were defined by electron-beam lithography and reactive ion etching in order to isolate the emission of a single QD. The excitation of the QDs and light collection strategies are similar to the ones described in Ref [2]. The histogram (fig.5(b)) obtained for the autocorrelation measurements under cw excitation was normalized and fitted using the correlation function calculated for a three level system[13] $g^{(2)}(\tau)=1-(1+a)\exp(-\tau/\tau_1)+a*\exp(-\tau/\tau_2)$. To account for the limited setup resolution, detection of uncorrelated photons and dark counts, we fitted the experimental data by convolving a gaussian time distribution (FWHM=200ps) with the correlation function corrected for noise[9]: $g_n^{(2)}(\tau)= 1+\rho^2(g^{(2)}(\tau)-1)$. The fit provides the values, $\rho=0.91$, $a=0.8$, $\tau_1=0.62$ns and $\tau_2=170.8$ns.

Acknowledgements: this work was supported by: Swiss National Foundation through the "Professeur borsier" and NCCR Quantum Photonics program, FP6 STREP "SINPHONIA" (contract number NMP4-CT-2005-16433), IP "QAP" (contract number 15848), NOE "ePIXnet" and the Italian MIUR-FIRB program.